\documentclass[fleqn,twoside]{article}
\usepackage{espcrc2}

\usepackage{graphicx}
\usepackage[figuresright]{rotating}

\newcommand{\AmS}{{\protect\the\textfont2
  A\kern-.1667em\lower.5ex\hbox{M}\kern-.125emS}}

\title{String theory: an update}

\author{J. de Boer\address{Institute for Theoretical Physics, University of Amsterdam\\
        Valckenierstraat 65, 1018 XE Amsterdam, The Netherlands}
        \thanks{{\tt jdeboer@science.uva.nl}} }

\begin{document}

\begin{abstract}
An overview of some of the developments in
string theory over the past two years is given, focusing on four topics:
realistic (standard model like) models from string theory,
geometric engineering and theories with fluxes, the gauge
theory-gravity correspondence, and time dependent backgrounds and
string theory. Plenary talk at ICHEP'02, Amsterdam, July 24-31,
2002. \vspace{1pc}
\end{abstract}

\maketitle

\section{INTRODUCTION}

In this talk I would like to discuss some of the developments that
took place in string theory over the last two years, after the
previous ICHEP meeting that took place in Osaka. Of course, it is
impossible to give a detailed account of everything that happened
in the field. In fact, it is not even clear precisely what the
field is. According to some people, ``anything that appears on
hep-th'' is a good first approximation to the term ``string
theory.'' Therefore, in order to have some degree of organization
and limit the material, I have selected four themes. Within each
of the four themes, there has been a significant amount of
activity over the past two years. The themes are (1) getting
realistic (standard model like) models from string theory, (2)
geometric engineering and theories with fluxes, (3) the gauge
theory-gravity correspondence, and (4) time dependent backgrounds
and string theory. I will discuss each of the themes, in this
order, in the four sections that follow.

Presenting such a general talk about string theory is a difficult
enterprize, and therefore I would like to start with a disclaimer.
Since the purpose of this talk is to give a flavor of what has
been happening in string theory over the past two years, many
interesting subjects will not be discussed. This by no means
implies that these subjects are less interesting or important,
there simply is too much material to cover and a selection has to
be made. In a similar vein, the list of references will be
extremely incomplete, and I apologize in advance for all omitted
references, and for all places where I did not give appropriate
credit.

An introduction to string theory and some general background
material may be found e.g. in the books \cite{background1}, review articles
\cite{background2}, popular articles \cite{background3} and the web page
\cite{background4}.

\section{REALISTIC MODELS FROM STRING THEORY}

The basic idea of string theory is to replace all point particles
in nature by small vibrating strings. Different ways in which the
string can vibrate manifest themselves as different particles.
Thus all particles become different excitations of one and the
same object. The typical size of a string is called the string
length $l_s$ and typically of the order of the Planck length, i.e.
$l_s \sim 10^{-33}$ cm. The main success of string theory is that
it can be made into a finite theory and quite generally contains
both gauge theories and gravity. It is therefore a candidate
theory that unifies all fundamental forces in nature. In
particular it provides us with a finite theory of quantum gravity.
As a field theory, gravity is not renormalizable, with the
divergences coming from the fact that it is a theory of
point-particles interacting at points in space-time. In string
theory, there is no longer a well-defined point where strings
interact, and this renders the theory finite.

It is not possible to directly see strings, as there are no
experiments that involve the large energies of order $10^{19}$ GeV
needed to resolve distances of order $10^{-33}$ cm. It is
therefore an important question how we might be able to
experimentally test string theory. Roughly, there are three
classes of possible experimental signatures.
\begin{enumerate}
\item
The very high energies that directly probe the string length scale
$l_s$ cannot be realized in an experiment done on earth, but do
certainly occur at extreme situations in the universe
\cite{cosm1}. In particular, the physics of the early universe and
the physics of black holes will crucially depend on a fundamental
theory of quantum gravity. Observations of the universe e.g. by
means of the cosmic microwave background, by means of (the still
to be observed) gravitational radiation, or by means of high
energy gamma rays may therefore provide experimental signatures of
string theory.
\item
At low energies, the standard model is very successful in
describing the interactions of fundamental particles. String
theory should be able to reproduce the standard model, and ideally
also constrain its free parameters. This would be quite
spectacular but has not yet been achieved. A less ambitious
project would be to obtain a reasonable supersymmetric extension
of the standard model, postponing the problem of supersymmetry
breaking. Evidence for the existence of such a supersymmetric
extension would for example be the discovery of a supersymmetric
partner of one of the known particles. Because supersymmetry is a
generic prediction of string theory, this would certainly be
enthusiastically received by the string theory community.
\item
One can contemplate other string theory models, such as ones
involving ``brane worlds'' or ``large extra dimensions,'' where
the string length is much larger than $10^{-33}$ cm. In these
scenarios, a bewildering set of possible experiments has been
proposed in the literature, including measurements of violations
of Newtonian gravity at length scales less than $0.2$ mm, and
missing energy signatures and black hole production at
accelerators; see also section~2.7.
\end{enumerate}
In addition, it is worth noticing that in many cases it is
difficult to say anything useful about each of the three points
above. That still leaves an important theoretical experiment,
namely
\begin{itemize}
\item[]
The theory should be self-consistent, and in particular one should
be able to embed it consistently in string theory.
\end{itemize}
Often, it is amazingly difficult to even perform this theoretical
experiment.

In this section, we want to mainly focus on the second point in
the list above, in particular we would like to discuss the
prospects for getting the standard model or one of its
supersymmetric extensions from string theory. In the last two
years, many new models have been proposed, and to explain how
these fit together, it is perhaps best to first go back in time.

Soon after its discovery, it was realized that superstring theory
is only consistent in ten dimensions. This is not a problem, since
as long as six of the ten dimensions are compact and very small,
the theory will at low energies look like a four-dimensional
theory. The details of the four-dimensional theory depend
crucially on the way in which the six extra dimensions are
compactified. In particular, the amount of supersymmetry that
remains in four dimensions depends on the choice of 6d geometry.
Among the several known ten-dimensional superstring theories,
there was only one that could be compactified in such a way that
only $N=1$ supersymmetry in four dimensions remained. This string
theory is called the heterotic string, and the six-dimensional
geometry that is needed is a so-called Calabi-Yau manifold
\cite{Candelas:en}.

In 1995 the picture changed dramatically in what is known as the
second superstring revolution \cite{secondrevolution}.
It was realized that the different
ten-dimensional string theories were not inequivalent theories,
but were all different weakly coupled limits of a single theory.
This is illustrated in figure~1, taken from \cite{polchinski}.

\begin{figure}
\begin{center}
\includegraphics[scale=0.6]{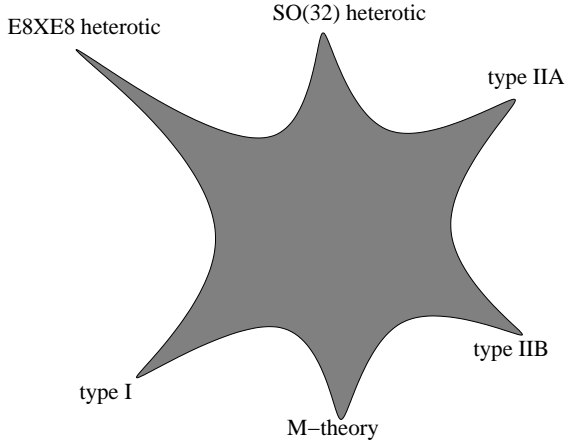}
\caption{string theory duality web}
\end{center}
\end{figure}

This picture illustrates the existence of one big theory, that
depends on many parameters. Depending on the choice of parameters,
there is at most one weakly coupled description of the theory, as
illustrated by the peaks in the picture. In the middle of the
picture, all coupling constants are of order one, and there is no
good weakly coupled description. Since the picture is connected,
it is possible to move from one theory to the other by changing
parameters. In particular, the theory one starts out with will
become strongly coupled by the time the other theory becomes
weakly coupled. These strong-weak coupling dualities play an
important role in string theory. A simple example is the duality
between M-theory and the type IIA superstring. Actually, M-theory
is a somewhat peculiar element in this picture, because it is not
really a string theory. It is a theory whose low-energy limit is
eleven-dimensional supergravity. Above eleven-dimensions,
supersymmetry necessarily involves fields of spin larger than two,
and such theories (with finitely many higher spin fields) do not
exist. There are indications that membranes play an important role
in M-theory, but if so their role is certainly different from that
of strings in string theory. In any case, the claim is that
M-theory compactified on a circle of radius $R$ yields the type
IIA superstring. This is certainly true at low energies, as one
can explicitly check that eleven-dimensional supergravity
compactified on a circle yields ten-dimensional type IIA
supergravity, which is the low-energy limit of type IIA
superstrings (hence the name). The string coupling constant
(usually denoted by $g_s$) that measures the strength of string
interactions is proportional to the radius $R$. For small $R$, the
coupling constant is small, and type IIA string theory is the
weakly coupled description. At large $R$, the coupling constant is
large, type IIA becomes strongly coupled, and a corresponding
weakly coupled description is in terms of M-theory.

The picture in figure~1 describes theories with a lot of
supersymmetry. One can try to make a similar ``duality web'' that
describes the situation of string theory compactified down to four
dimensions with $N=1$ supersymmetry remaining. A caricature of
such a picture is given in figure~2. This figure is necessarily
incomplete, since there are many more possible weakly coupled
possibilities than are shown in figure~2. In addition, it is not
even known for sure whether the picture should be connected or in
reality consists of several disconnected components. Nevertheless,
we will use this picture as a guiding principle to discuss some of
the new ways in which one can obtain $N=1$ supersymmetric theories
in four dimensions (theories with fluxes will be discussed in
section~3). Ideally, one would like to study models with no
supersymmetry at all (like the standard model), but supersymmetry
breaking remains an interesting and difficult problem in string
theory. It is possible to explicitly break supersymmetry, but then
it is often difficult to examine whether the theory is stable or
not. Therefore we will restrict attention to $N=1$ supersymmetric
theories in what follows.

\begin{figure}
\begin{center}
\includegraphics[scale=0.53]{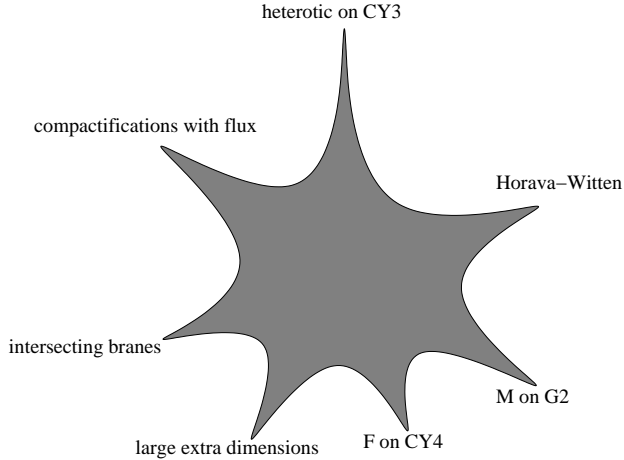}
\caption{$N=1,d=4$ string theory duality web}
\end{center}
\end{figure}

\subsection{heterotic on CY${}_3$}

As we mentioned before, an $N=1$ theory in four dimensions was
first obtained by compactifying the heterotic string on a
Calabi-Yau manifold. The 3 in CY${}_3$ refers to the number of
complex dimensions of the Calabi-Yau manifold. These models have
many appealing properties. They can naturally accommodate GUT
groups such as $SU(5)$, $SO(10)$ and $E_6$, it is not too
difficult to get three generations of chiral fermions in four
dimensions, gauge coupling unification is quite natural, etc. The
GUT scale is two or three orders of magnitude smaller than the
Planck scale. There is a large number of models that contain the
standard model fields, but no model that precisely gives the MSSM
(minimally supersymmetric standard model). A generic problem of
these models is the large number of massless scalar fields that
they posses; these parameterize the different shapes and sizes of
the Calabi-Yau manifold. It is possible that a potential for these
scalar fields is dynamically generated once supersymmetry is
broken, but unfortunately we do not know the precise mechanism for
supersymmetry breaking. For more discussion of low energy
phenomenology in string theory, see e.g.
\cite{heteroticdiscussion}.

\subsection{Horava-Witten}

The Horava-Witten scenario \cite{horavawitten} is based on a
strong-weak coupling duality, as figure~2 suggests. The duality in
question is that between the heterotic string on one side, and
M-theory compactified on an interval on the other side. Although
it may sound strange to compactify a theory on an interval, there
is nothing wrong with this. It is similar to doing field theory in
a finite box, though one has to be careful to choose appropriate
boundary conditions for all fields at the endpoints of the
interval. It turns out that consistency requires the introduction
of additional boundary degrees of freedom that live only at the
endpoints of the interval. These boundary degrees of freedom can
freely propagate in the remaining ten dimensions, but are stuck in
the eleventh dimension. In particular, they can freely propagate
in the four dimensions that remain once the theory is further
compactified on a Calabi-Yau space, and to the low-energy observer
they look like conventional four-dimensional degrees of freedom.
Sometimes one says that there are ``end of the world branes''
located at the endpoints of the interval that carry additional
degrees of freedom.

The duality with the heterotic string is a strong-weak coupling
duality in the sense that for a small interval, the heterotic
string coupling is small and that is the weakly coupled
description, whereas for large string coupling the interval
becomes larger and the M-theory point of view is more appropriate.

The phenomenology of these models is quite rich. It includes all
heterotic compactifications, but there are additional
possibilities. One can tune parameters in such a way that the GUT
scale and the Planck scale coincide. The study of the
Horava-Witten models involves some rather heavy geometric
machinery, but there are indications that models with three
generations and gauge group $SU(3)\times SU(2) \times U(1)$ can be
obtained. For further discussion, see \cite{horavawittenlist} and
references therein.

\subsection{M on G${}_2$}

The only theory whose low-energy limit is eleven dimensional
supergravity is M-theory. In order to obtain a four-dimensional
theory, seven dimensions need to be compactified. In case of the
heterotic string, we needed a special type of manifold, namely a
Calabi-Yau manifold, in order to have unbroken $N=1$ supersymmetry
in four dimensions. Similarly, in the case of M-theory, we need a
special type of seven manifold, so-called manifolds of $G_2$
holonomy, to have unbroken $N=1$ supersymmetry in four
dimensions\footnote{The $G_2$ refers to the exceptional Lie group
$G_2$. It can be defined as the subgroup of $SO(7)$ that preserves
a single spinor in the eight-dimensional spinor representation of
$SO(7)$. This spinor is responsible for the unbroken $N=1$
supersymmetry in four dimensions.} . The geometry of such
seven-dimensional manifolds is more complicated and less
well-understood than the geometry of Calabi-Yau manifolds. Several
examples of compact seven manifolds of $G_2$ holonomy are known
\cite{compex}. One also often studies non-compact manifolds of
$G_2$ holonomy. These are not useful to build a realistic theory
in four dimensions, because on non-compact manifolds particles can
have arbitrarily small momenta in the non-compact directions.
These manifest themselves as a continuum of particles in four
dimensions, which clearly is not a very realistic feature.
However, non-compact manifolds provide useful toymodels to understand the
structure of singularities. Singularities in $G_2$ manifolds are
of crucial importance because they are the only source of 
chiral fermions in such compactifications,
a fact that further complicates matters and makes
it difficult to obtain realistic models; for more
discussion see e.g. \cite{g2list}.

\subsection{F on CY${}_4$}

In this title\, F refers to F-theory, introduced in \cite{vafa}.
F-theory is a twelve-dimensional theory, but in contrast to
M-theory it is not generally covariant in twelve-dimensions. One
may even wonder what is really meant by the word theory in
``twelve-dimensional theory.'' If nothing else, F-theory is a
convenient way to describe in purely geometric terms certain
strongly coupled type IIB superstring compactifications. To obtain
a four-dimensional theory with $N=1$ supersymmetry, we need to
compactify F-theory on an eight-dimensional (or four
complex-dimensional) Calabi-Yau manifold. Though the number of
possible compactifications of F-theory is very large, we are not
aware of any specific phenomenologically appealing model that can
only be realized in this context.

\subsection{duality between branes and geometry}

\label{branesec}

Before continuing with our discussion of the remaining
possibilities given in figure~2, we would like to briefly discuss
the notion of branes, and how theories with branes can be dual to
theories without branes. More information on this subject can be
found in most of the reviews listed under
\cite{background1,background2,background3,background4}.

D-branes are certain extended objects in string theory, that were
introduced by Polchinski in \cite{polchinski2}. They are labelled
by the number of dimensions of the object, so that a D0 brane is
like a particle, a D1 brane is like a string, a D2 brane is like a
membrane, etc. There are two ways to think about D-branes. On the
one hand, they are solitonic solutions of the equations of motion
of low-energy closed string theory. On the other hand, they are
objects in open string theory with the property that open strings
can end on them. Open strings have a finite tension, and their
center of mass cannot be taken arbitrarily far away from the
D-brane. As a consequence, the degrees of freedom of the open
string can effectively only propagate in a direction parallel to
the brane: one says that they are confined to the brane, or that
they live on the brane. The open string spectrum can be reproduced
directly from the soliton in the closed string description via a
collective coordinate quantization.

As a very crude analogy, one can think about two ways to describe
a monopole. On the one hand, one can think of it as the 't
Hooft-Polyakov monopole, in which case it is an extended soliton
solution of the Yang-Mills-Higgs equations of motion. On the other
hand, one can view a monopole as a point particle, on which
magnetic field lines can end. Both descriptions have their
advantages, as do the open and closed string descriptions of
D-branes.

D-branes play a crucial role in string theory and in particular in
string dualities. For example, type IIB string theory has a
strong-weak coupling duality that inverts the the type IIB string
coupling constant, $g_s \leftrightarrow 1/g_s$. Under this
duality, the roles of fundamental strings and D1 branes are
interchanged (both are one-dimensional objects). Therefore, D
branes appear to be as fundamental as strings themselves. There is
a close analogy between this duality and the electic-magnetic
duality of the Maxwell equations, with strings playing the role of
electric charges, and D-branes the role of magnetic charges.

Interestingly, D branes can not only be dual to other branes and
strings, but also to non-trivial geometries without branes. To
explain this peculiar statement, we will use the duality between
M-theory compactified on a circle, and type IIA string theory.
M-theory at low energies was described by eleven-dimensional
supergravity. When compactifying this on a circle, the eleven
dimensional metric $g_{mn}$ produces a gauge field $A_{\mu}$ in
ten dimensions via $A_{\mu}=g_{11\mu}$. Now type IIA string theory
has objects that are charged with respect to this gauge field. In
four dimensions, charges can be measured by looking at the flux of
the electric or magnetic field through a two-sphere $S^2$
surrounding the charge. In other words, we compute $\int_{S^2} F$,
with $F_{\mu\nu}=\partial\mu A_{\nu}-\partial_{\nu} A_{\mu}$ the
field strength. A point particle in ten dimensions on the other
hand does not carry a charge under such an $F$, because we cannot
surround it by a two-sphere. We can surround it by an
eight-sphere, and if we would have an eight-index field strength
we could integrate that over the eight sphere and use that to
define a charge. However, besides string excitations, type IIA
string theory also has D-branes, and the list of possible D-branes
includes a D6-brane. A D6 brane can be surrounded by a two sphere
in ten dimensions. In general, in $d$ dimensions a $Dp$ brane can
be surrounded by a $d-2-p$ sphere. It is easy to verify this
directly in four and lower dimensions. Thus, a D6 brane can carry
a charge with respect to the gauge field $A_{\mu}$ of
ten-dimensional type IIA supergravity, measured by the flux
through the two-sphere. One can show that it indeed does carry
such a charge. In other words, a D6-brane is surrounded by a
non-trivial configurations of the gauge field $A_{\mu}$.

What does a D6-brane look like in eleven-dimensions? In eleven
dimensions, the gauge field $A_{\mu}$ was part of the
eleven-dimensional metric. Therefore, from the eleven-dimensional
perspective, a D6-brane is surrounded by a non-trivial geometry.
Since the gauge field is no longer there, the D6 brane is no
longer charged under anything. It has become a purely geometrical
object. One can show that the geometry is the product of
seven-dimensional Minkowski space and an Eguchi-Hanson space,
which is a gravitational instanton in four dimensions.

\subsection{intersecting branes}

D-branes can also be used to construct many new string theories
with $N=1$ supersymmetry in four dimensions. The idea is to take
any existing string theory, and to add branes to it. If we are
interested in keeping Poincar\'e invariance in four dimensions,
the branes need to have at least three space dimensions, so that
they can completely fill the four noncompact dimensions where the
low-energy observer lives. In the remaining compact six dimensions
(seven for M-theory) the branes can have any shape and size that
is compatible with supersymmetry. At low energies, besides the
fields that we get from the string theory we started out with, we
also get degrees of freedom from the various branes. In their
presence, the theory necessarily contains open strings that start
and end on the branes, and these give rise to additional degrees
of freedom in four dimensions. The new degrees of freedom never
contain gravity, but typically provide gauge fields and matter
fields. Gravity still has its origin in the original closed string
theory. The different origin of gauge fields and gravity plays an
important role in the large extra dimension scenario in the next
subsection.

There are many possible gauge groups and matter fields that one
can get from branes. For example, nonabelian gauge symmetries can
be obtained by putting several branes on top of each other. Open
strings can stretch from each of the branes in the stack to any of
the other branes. There are $N\times N$ different open strings and
these transform in the adjoint representation of $U(N)$, where $N$
is the number of branes. Their degrees of freedom include a
nonabelian $U(N)$ gauge field in four dimensions. A more
complicated possibility is to have branes that intersect in the
six (or seven) compact dimensions. A simplified picture of such
intersections is two orthogonal planes that intersect along a
line. Along the line, new degrees of freedom are localized due to
the open strings that stretch from one of the planes to the other.
A similar story applies in higher dimensions, where the new fields
that are confined to the intersections often give chiral matter in
$d=4$.

One of the simplest concrete models with intersecting branes is to
start with the type IIA string theory compactified on a six-torus,
i.e. each of the six compact dimensions is taken to be a circle.
One can take D6 branes and add these to the theory. Since three of
the six dimensions of the brane need to be along the three
noncompact dimensions of our world, the remaining three have to be
put on the six-torus; one can for example select three of the six
compact directions, and demand that the brane fills those three.
By selecting different sets of three of the six compact
directions, we can put different sets of D6 branes on the
six-torus, and this collection of intersecting D6 branes gives
rise to interesting theories in four dimensions.

There are several advantages to such an approach. It is possible
to obtain $N=1$ theories in four dimensions without having to rely
on the complicated geometry of Calabi-Yau manifolds, but one can
work with things as simple as a six-torus. The matter content is
relatively easy to control, simply by analyzing the open string
that can stretch between the different branes. In this way, the
chiral spectrum of the standard model and its minimal
supersymmetric extension have been obtained. It is even possible
to choose brane configurations that explicitly break
supersymmetry, but the stability of these configurations is not
clear. Another feature of intersecting brane models is the
stability of the proton, because baryon number becomes a gauge
symmetry.

In the past two years extensive work on these models has been
done. We refer the reader to \cite{longlist} and references
therein for further information and details.

\subsection{large extra dimensions}

The large extra dimension scenario \cite{ledoriginal} is strictly
speaking part of the intersecting brane story, but it is worth to
point out how this idea fits in the previous discussion.

As we explained, D-branes give rise to extra degrees of freedom,
typically gauge fields and matter. Gravity still comes from the
closed string sector. Now if we imagine that all gauge fields and
matter fields of the standard model originate from a certain set
of D-branes, then in some sense we are living on these branes. The
only way to probe the compact directions transversal to the branes
is by means of gravity. Since gravity has been poorly tested on
length scales less than 0.2 mm \cite{newton}, there is no
immediate contradiction with experimental data if we take the
compact directions transversal to the brane to be quite large, up
to sizes of the order of $\sim$0.1 mm (but see \cite{newton2}).
The standard model interactions are confined to the brane and do
not directly see the extra dimensions. This asymmetric treatment
of gravity and gauge interactions is the basic idea behind the
large extra dimension scenario. The ten dimensional string length
can be much larger than the $10^{-33}$ mm mentioned before, while
the Newton constant in four dimensions remains as small as it is,
and therefore these models are sometimes called models with a low
string scale. For more discussion, references, and experimental
implications, see e.g. \cite{largeextra}.

\subsection{summary}

Though all different possibilities we described look quite
different, they are all part of a huge duality web. This is an
important lesson. Compactifications with branes, as well as the
large extra dimension scenario are not ideas that are completely
disjoint from the original heterotic string on a Calabi-Yau
manifold. They merely represent different strongly coupled
versions of each other. In particular, branes provide a generic
ingredient in string theory compactifications.

Nevertheless, the heterotic string on a Calabi-Yau manifold (and
the Horava-Witten scenario) are still the most promising. They are
the only models that naturally incorporate gauge coupling
unification. In the other models gauge coupling unification is
more artificial.

A problem shared by virtually all models is the existence of
additional low energy degrees of freedom beyond those of a
reasonable supersymmetric extension of the standard model. This
can perhaps be resolved once the mechanism of supersymmetry
breaking is better understood.

Despite this, there are several possible experimental signatures
that do not depend very much on the details of the model, and the
search for such signatures, both theoretically and experimentally,
is clearly an important and urgent problem.

\section{GEOMETRIC ENGINEERING AND THEORIES WITH FLUXES}

We postponed the discussion of theories with fluxes to this
separate section, because there has been quite a lot of recent activity
in this area.

First we need to explain the terms ``theories with fluxes'' and
``geometric engineering.'' The starting point is the type II
string compactified on a Calabi-Yau manifold. Recall that the
heterotic string, when compactified on a Calabi-Yau manifold, gave
rise to an $N=1$ theory in four dimensions. Type II strings have
twice as many supersymmetry as the heterotic string, and
therefore yield four dimensional theories with $N=2$
supersymmetry instead.

Geometric engineering, a term which appears to have been introduced in
\cite{geomen}, refers to the process of constructing a
suitable Calabi-Yau geometry that produces at low energies an a
priori given $N=2$ gauge theory in four dimensions. It turns out
that many gauge theories can be geometrically engineered. Typical
gauge theories that appear are of ``quiver'' or ``moose'' type.
Such theories have gauge groups that are a product of different
factors, and matter fields that transform non-trivially under one
or two of the gauge groups. Matter fields transforming under three
or more of the gauge groups do not appear. Interestingly, this is
related to the fact that strings have only two endpoints.

It is possible to break these $N=2$ theories to $N=1$ by adding
branes, adding fluxes, or both. The dualities between these
possibilities have been explored in great detail by many authors.

We already discussed the possibility of adding branes. Adding
fluxes refers to the following \cite{flux}.
Type II string theory has several
massless tensor fields besides the graviton. For example, type IIA
string theory also contains a gauge field and a three-index antisymmetric tensor
field. This gauge field was crucial in the relation between type IIA
string theory and M-theory. Now suppose that the Calabi-Yau
manifold looks like, say, a two-sphere. Then we can turn on a
non-zero field strength for the gauge field, in such a way that
there is a non-zero magnetic flux through the two-sphere. In other
words, the integral of the field strength over the two-sphere is
non-zero. The gauge field configuration is similar to the gauge
fields that surround a magnetically charged particle sitting in
the middle of a two-sphere. Of course, here the $S^2$ is empty,
and in particular there is nothing in the
interior of the two-sphere and in particular there no actual
magnetically charged particle that sits there.

More generally, the non-trivial geometry of the Calabi-Yau
manifold can be used to give simultaneous expectation values to
several of the field strengths of the massless tensor fields that
appear in type II string theory.

Such fluxes give rise to a non-zero energy density, and this energy will
back-react on the geometry. Thus, after the fluxes are turned on, the
geometry of the space will no longer be that of a Calabi-Yau
space. The corrections to the Calabi-Yau geometry are supressed by
one over the volume, so as long as the size of the Calabi-Yau is
sufficiently large this is not a problem.

Type II string theories with fluxes have many interesting
properties:

\begin{enumerate}

\item They have a rich duality structure. The full mathematical structure of
fluxes and branes remains to be uncovered, but it is clear that
very advanced mathematics will play an important role, see e.g.
\cite{fluxdual}.

\item The gauge kinetic terms and superpotential of the low-energy
effective field theory in four dimensions can often be computed
exactly using topological string theory. Topological string theory
is a reduced version of ordinary string theory, where many of the
degrees of freedom have disappeared. In particular, as the name
topological suggest, the metric is no longer a degree of freedom,
and in fact there are no local degrees of freedom at all. Perhaps
the simplest example of a topological theory is Chern-Simons
theory in $2+1$ dimensions, which does not depend on any metric in
$2+1$ dimensions, and the only observables of the theory are
Wilson lines. Interestingly, certain correlation functions in
topological string theory are identical to those in the full
string theory, and since topological string theory is so much
simpler this has allowed for the explicit computation of a whole
class of correlation functions. The gauge kinetic terms and
superpotential in four dimensions belong to this class. Besides
this important physical application, topological string theory
calculations have also yielded many new results in mathematics.
For more, see \cite{topstringlist}.

\item As we stressed in section~2, many string
compactifications yield additional massless degrees of freedom at
low energies beyond those of a (reasonable supersymmetric
extension of the) standard model. In particular, there are often
additional massless scalar fields, called moduli, that
parameterize the shape and size of the Calabi-Yau manifold. When
we compactify the type II string, and also include some fluxes, a
superpotential is generated that can explicitly be computed
\cite{flux}. This superpotential gives rise to mass terms for many
of the moduli fields, which can thereby be removed from the
low-energy effective field theory. It is much easier to analyze
the superpotential in these models than for example in the
heterotic string compactified on a Calabi-Yau manifold, and this
is a significant advantage.

\item Another generic feature of compactifications with fluxes is the
appearance of ``warped'' compactifications. In usual Kaluza-Klein
like compactifications, the full ten-dimensional background is a
direct product of a four-dimensional Minkowski space-time and a
six-dimensional Calabi-Yau manifold. Four-dimensional Poincar\'e
invariance can be preserved if we allow for a more general
``warped'' setup, where the metric on Minkowski space is rescaled
by an overall factor that depends non-trivially on the coordinates
on the Calabi-Yau manifold. For example, suppose that our world is
a circle and the total space is a cone. The size of the circle
varies depending on where we are on the cone, but rotational
invariance remains unbroken. In warped compactifications, there is
a relation between the energy scale in four dimensions and the
additional coordinates of the Calabi-Yau manifold. Moving along
the Calabi-Yau manifold changes the overall factor of the metric
on Minkowski space and therefore also the energy. The close
relation between extra dimensions and the energy scale in four
dimensions also appears in the duality between gauge theories and
gravity that we discuss in the next section. See
\cite{largeextra,flux} for more discussion and literature.

\item The various dualities that have been established in the
context of type II strings with branes and/or fluxes have provided
topological versions of the gauge theory-gravity correspondence
which we discuss in the next section, with interesting
mathematical and physical applications \cite{gopakumarvafalist}.

\end{enumerate}

One such recent application is the following \cite{dijkgraafvafa}:
Consider a supersymmetric $N=1$ gauge theory in four dimensions
with classical superpotential $W_0$. Assume that $W_0$ has an
isolated critical point where the gauge group is broken to a
product of $U(N_i)$ factors. Pure $U(N_i)$ $N=1$ gauge theories
exhibit confinement at low energies which is signaled by a
condensation of the gluino condensate field $S_i={\rm tr}(\lambda
\lambda)$, where $\lambda$ is the gluino, the superpartner of the
$U(N_i)$ gauge field. The gluino condensate in pure $N=1$ theories
is described by an effective quantum superpotential $W_{eff}(S_i)$
known as the Veneziano-Yankielowicz superpotential \cite{vysuper}.
Returning to the case where a classical superpotential $W_0$
breaks $U(N)$ to a product of $U(N_i)$ factors, one may wonder
what the analogue of the Veneziano-Yankielowicz superpotential for
such a theory is. In \cite{dijkgraafvafa} it is conjectured that
this quantum superpotential can be computed by simply summing
planar diagrams in $0+0$ dimensional system, namely the matrix
theory with action $W_0$. In many examples this can be proven
using various string dualities and properties of topological
strings. A complete proof directly on the level of field theory
has been given in a special case in \cite{proof1} and in much more
generality in \cite{proof2,proof3}. There has also been
independent progress in all-order field theoretic instanton
calculations in $d=4$, $N=2$ gauge theories \cite{nekrassov},
which may be related.

\section{GAUGE THEORY-GRAVITY CORRESPONDENCE}

The gauge theory-gravity correspondence refers to an amazing
equivalence between certain theories with gravity, and certain
theories without gravity. One particular example of such a
duality, as originally conjectured by Maldacena \cite{maldacena},
is the exact equivalence between type IIB string theory
compactified on $AdS_5 \times S^5$, and four-dimensional $N=4$
supersymmetric Yang-Mills theory. The abbreviation $AdS_5$ refers
to an anti-de Sitter space in five dimensions, $S^5$ refers to a
five-dimensional sphere. Anti-de Sitter spaces are maximally
symmetric solutions of the Einstein equations with a negative
cosmological constant. The large symmetry group of 5d anti-de
Sitter space matches precisely with the group of conformal
symmetries of the $N=4$ super Yang-Mills theory, which for a long
time has been known to be conformally invariant. In view of this,
the gauge theory-gravity correspondence is often referred to as
the AdS/CFT duality, where CFT stands for conformal field theory.

Anti-de Sitter space can be roughly thought of as a product of
four-dimensional Minkowski space times an extra radial coordinate.
The metric on Minkowski space is however multiplied by an
exponential function of the radial coordinate, and Anti-de Sitter
space is therefore an example of a warped space: in a suitable
local coordinate system, $ds^2=dr^2 + e^{2r}
(\eta_{\mu\nu}dx^{\mu} dx^{\nu})$. The limit where the radial
coordinate goes to infinity and the exponential factor blows up is
called the boundary of Anti-de Sitter space. This boundary is the
place where the dual field theory lives. One can indeed verify
that string theory excitations in anti-de Sitter space extend all
the way to the boundary. In this way one obtains a map from string
theory states to states in the field theory living on the
boundary.

Is is very hard to directly prove the equivalence between type IIB
string theory on $AdS_5 \times S^5$, and four-dimensional $N=4$
super Yang-Mills theory. For one, we do not have a good definition
of non-perturbative type IIB string theory. Even at string tree
level, we do not (yet) know how to solve the theory completely.
From this perspective, it is perhaps better to view $N=4$ super
Yang-Mills theory as the definition of non-perturbative type IIB
string theory on the $AdS_5 \times S^5$ background.

A weaker form of the AdS/CFT corespondence is obtained by
restricting to low-energies on the string theory side. At
low-energies, type IIB string theory on $AdS_5 \times S^5$ reduces
to type IIB supergravity on $AdS_5 \times S^5$. The corresponding
limit on the gauge theory side is one where both $N$ and $g_{YM}^2
N$ become large, where $N$ is the rank of the $U(N)$ gauge group
of the $N=4$ supersymmetric gauge theory (not to be confused with
the $N$ appearing in $N=4$), and $g^2_{YM}$ is the gauge coupling
constant. The equivalence between type IIB supergravity on $AdS_5
\times S^5$ and $N=4$ gauge theory in the large $N$, large
$g^2_{YM} N$ limit has been very well tested by now.

The AdS/CFT correspondence is related to two deep ideas in
physics.

The first of these is the idea that large $N$ gauge theory is
equivalent to a string theory \cite{thooft}. The perturbative
expansion of a large $N$ gauge theory in $1/N$ and $g^2_{YM} N$
has the form of a string loop expansion, with the string coupling
$g_2$ equal to $1/N$. Through some peculiar and not completely
understood mechanism, Feynman diagrams of the gauge theory are
turned into surfaces that represent interacting strings (but see
\cite{oogurivafa}). Apparently, this is precisely what happens in
the AdS/CFT correspondence.

The second is the idea of holography \cite{thooft2,susskind}. This
idea has its origin in the study of the thermodynamics of black
holes. It was shown by Bekenstein and Hawking \cite{bekhaw} that
black holes can be viewed as thermodynamical systems with a
temperature and an entropy. The temperature is directly related to
the black body radiation emitted by the black hole, whereas the
entropy is given by $S=A/4G$, with $G$ the Newton constant and $A$
the area of the horizon of the black hole. With these definitions,
Einstein's equations of general relativity are consistent with the
laws of thermodynamics. Since in statistical physics entropy is a
measure for the number of degrees of freedom of a theory, it is
rather surprising to see that the entropy of a black hole is
proportional to the area of the horizon. If gravity would behave
like a local field theory, one would have expected an entropy
proportional to the volume. A consistent picture is reached if
gravity in $d$ dimensions is somehow equivalent to a local field
theory in $d-1$ dimensions instead. Both have an entropy
proportional to the area in $d$ dimensions, which is the same as
the volume in $d-1$ dimensions. The analogy of this situation to
that of an hologram, which stores all information of a 3d image in
a 2d picture, led to the name holography. The AdS/CFT
correspondence is holographic, because it states that quantum
gravity in five dimensions (forgetting the compact five sphere) is
equivalent to a local field theory in four dimensions.

One of the questions that the AdS/CFT immediately raises is that
of the interpretation of the extra fifth dimension in the field
theory. It turns out that it is closely related to the energy
scale. From the 5d gravitational point of view, low-energy
processes in field theory stay close to the boundary of AdS,
whereas high-energy processes penetrate deeper in the interior
\cite{wisu}. One can even show that the invariance under 5d
general coordinate transformations implies the Callan-Symanzik
renormalization group equations in the field theory \cite{bvv}.
Thus, from the 5d point of view, the renormalization group is on
an equal footing with 4d Poincar\'e invariance.

The AdS/CFT correspondence is also an example of a weak/strong
coupling duality. Depending on the choice of parameters, either
AdS or the CFT is a weakly coupled description of the system, but
never both at the same time. Gauge theory is a good description
for small $g^2_{YM} N$ and small $g^2_{YM}$, whereas string theory
is good for large $g^2_{YM}N$ and small $g^2_{YM}$. Therefore, the
AdS/CFT correspondence can be applied in two directions. We can
use string theory to learn about gauge theory, and we can use
gauge theory to learn about string theory.

One of the most difficult and unsolved problems in the AdS/CFT
correspondence is to reconstruct 5d local gravitational physics
directly from the dual 4d field theory point of view. In
particular, we would like to know in what way the local
gravitational description breaks down. Does such a breakdown occur
in a local way at the Planck length, are in a non-local way at
much larger length scales? The AdS/CFT correspondence seems to
prefer the second answer, which is also the answer that may
provide a resolution to the black hole information paradox. This
paradox is based on the fact that semiclassically, everything that
falls into a black hole is converted into purely thermal
radiation, with no memory of the object that fell in except for
its mass and perhaps a few other quantum numbers. Such a process
contradicts the usual rules of quantum mechanics, and we can
either give up on quantum mechanics or give up on the
semiclassical approximation to quantum gravity; AdS/CFT prefers
the latter.

The emergence of a concrete duality between a theory with gravity
and a theory without gravity is one of the most important results
of string theory. Below, we summarize some of the developments in
this area over the past two years. For more, see the reviews
\cite{adscftreview}

\subsection{high energy scattering/deep inelastic
scattering}

At first sight, the AdS/CFT correspondence, or any duality between
string theory and gauge theory, seems at odds with the known fact
that the scattering of glueballs at high energies is hard, whereas
string scattering at high energies is soft, due to their extended
nature. The resolution sits in the fact that AdS is a warped
space. When an object moves away from the boundary of AdS, its
size is exponentially reduced. Very high energy processes in the
gauge theory are described by strings which propagate a long
distance from the boundary of AdS before they interact. By that
time, the size of the strings has been exponentially reduced, and
this compensates for the softness of string interactions to make
it into a hard process in the gauge theory
\cite{polchinskistrassler}. Besides such effects, which are due to
the geometric warping of AdS, other gauge theory processes
crucially involve strong gravity physics like black hole formation
\cite{giddings}. It is also possible to study the physics of deep
inelastic scattering and the parton model from the AdS/CFT point
of view \cite{ps2}.

\subsection{towards a QCD string?}

A more involved version of the AdS/CFT correspondence is the one
given in \cite{klebanovstrassler}. It was discovered by studying
branes stuck in singularities  in string theory. The gauge theory
that appears is an $N=1$ theory in four dimensions with gauge
group $SU(N) \times SU(N+M)$. There are two chiral superfields
$A_i$ in the $(N,\overline{N+M})$ representation of the gauge
group, and two chiral superfields $B_i$ in the $(\overline{N},
N+M)$ representation. In addition, there is a nontrivial
superpotential of the form $W\sim \epsilon^{ij} \epsilon^{kl} {\rm
tr}(A_i B_k A_j B_l)$.

This field theory has a remarkable property: it has running gauge
couplings, but does not become free at either low or high
energies. The gauge coupling becomes strong either way. Strongly
coupled $N=1$ theories in four dimensions often admit a dual
weakly coupled description, a duality known as Seiberg duality
\cite{seibdual}. The same is true here: both at low energies and
at high energies there exist dual descriptions. However, these
dual descriptions have the same problem: they are not weakly
coupled at either low or high energies. Again, they admit suitable
dual descriptions. The full picture that emerges is that of an
infinite ``cascade'' of gauge theories, that continues
indefinitely at high energies, with an ever increasing rank of the
gauge group, but terminates at low energies once e.g. the rank of
one of the gauge groups becomes one. At that point, the gauge
theory becomes confining. Strictly speaking we need an infinite
amount of fine tuning of irrelevant operators to obtain this
infinite cascade, but quite remarkable, the dual description of
this gauge theory quite naturally sees the same cascade. The ranks
$N$ and $M$ of the gauge group become non-trivial functions of the
radial coordinate of the dual AdS-like geometry. This also
confirms once more the interpretation of the extra fifth dimension
in the AdS/CFT correspondence as an energy scale in the field
theory.

The AdS-like geometry that is dual to this infinite
cascade has several nice features. String theory on
this background exhibits (i) confinement, (ii)
glueballs and baryons with a mass scale that emerges
through dimensional transmutation, exactly as in the
gauge theory, (iii) gluino condensates that break the
${\bf Z}_{2M}$ chiral symmetry to ${\bf Z}_2$,
and
(iv)  domain walls separating different vacua.

The gauge theory at low energies reduces to a pure $N=1$
supersymmetric Yang-Mills theory. Does the dual geometry therefore
provide a dual string theory for pure supersymmetric Yang-Mills
theory, the long sought for QCD string? Not really, because  it
has new degrees of freedom beyond those of the field theory that
appear at $\Lambda_{QCD}$. This is a generic problem in trying to
find weakly coupled string theory descriptions of gauge theories.
To decouple the additional degrees of freedom, we need to make the
curvature of the AdS-like geometry large, while keeping the string
coupling $g_s$ small. String theory in a strongly curved
background is described by a strongly coupled $1+1$ dimensional
field theory. The structure of the sigma models relevant for the
AdS/CFT correspondence is not very well understood, but there has
been progress in this direction recently (see \cite{berkovits} and
references therein) , and the prospect of finding a string theory
dual of QCD remains an exciting possibility.

\subsection{other string effects in gauge theories:
large quantum numbers and pp-waves}

Instead of trying to find a precise string theory dual description
of pure $N=1$ supersymmetric Yang-Mills theory, it is also
interesting to look for more qualitative stringy behavior in gauge
theories.

One place to find such behavior is to look at states with a large
scaling dimension proportional to $N$, the rank of the gauge
group. Many gauge theories have baryons with such scaling
dimensions, and it turns out that they are not described by
strings but by branes in the dual geometrical description
\cite{baryon}. Thus, it is also possible to discover branes in
gauge theory.

A related example is to consider operators with a large spin $s$,
like for example ${\rm tr}(\Phi D_{\mu_1} \ldots D_{\mu_s} \Phi)$,
where $\Phi$ is some field that transforms in the adjoint
representation of $U(N)$. Such operators correspond to folded
rotating closed strings in the dual geometrical description. One
can compute the scaling dimension of these operators both in the
field theory and in the dual geometrical description. This
confirms the equivalence between the two, as one finds in both
cases that it behaves like $s+\log s$ \cite{gkpgonzalezetc}.

A more complete way to recover string theory from a gauge theory
has been described in \cite{bmn}. The idea is to take a particular
scaling limit of the AdS/CFT correspondence. This scaling limit,
when applied to the AdS geometry, yields a different geometry
known as a ``pp-wave''. In fact, many geometries admit scaling
limits in which they reduce to pp-waves, as originally shown by
Penrose \cite{penrose}. String theory on the pp-wave, in the
absence of string interactions, can be exactly solved, and in
particular the free string spectrum can be obtained.

On the field theory side, the same limit can be taken. In this
limit only a subset of the operators of the full $N=4$ super
Yang-Mills theory survive, namely those for which the scaling
dimension $\Delta$ and a certain global $U(1)$ quantum number $J$
have the property that $\Delta+J$ scales as $N^{1/2}$, while
$\Delta-J$ is kept finite, as one takes $N\rightarrow \infty$.

Interestingly, this set of operators is in one-to-one
correspondence with the set of free string states. This is the
first example where a complete string spectrum has been obtained
from a gauge theory, albeit in a special scaling limit.

The ground state of the string theory (in light-cone quantization)
is described in the gauge theory by the operator ${\rm tr}(Z^J)$,
where $Z$ is a complex scalar field in the adjoint representation
of the gauge group with charge $J=+1$ under the distinguished
global $U(1)$ symmetry. The simplest excited states of the string
are operators of the form $\sum_{i} a_i {\rm tr}(Z^i \Phi
Z^{J-i})$ where the $a_i$ are phases. The string appears from this
point of view as composed of ``string bits.'' The string bits are
the operators $Z$, and the string is composed of a string of $J$
bits. Exciting the string amounts to introducing impurities like
$\Phi$ that are distributed with phases (i.e. a discrete momentum)
along the chain of $Z$'s.

A similar discretized picture of string theory can be obtained in
several other gauge theories as well, see e.g. \cite{pp1}. It is
also presently being investigated whether one can correctly
recover string interactions, or even the full string field theory,
from the gauge theory \cite{pp2}.

\section{TIME DEPENDENT BACKGROUNDS AND STRING THEORY}

The celebrated type IA supernovae measurements \cite{supern} of a
few years ago that showed the existence of a small but nonzero
positive cosmological constant are partly responsible for a
renewed interest in time-dependent backgrounds in string theory.
It is much more difficult to obtain a small nonzero cosmological
constant than a cosmological constant that is strictly zero. The
latter can arise for example due to an underlying symmetry. A
small positive cosmological constant on the other hand introduces
a new scale in the theory, and this has to be put in by hand or it
should be explained by some unknown mechanism in terms of the
existing length scales. Although various mechanisms have been
proposed in the literature, no completely satisfactory explanation
has been given. Therefore, most attention has recently been
focused on trying to understand time-dependent backgrounds and
cosmology at a more conceptual level. In particular, many
cosmological scenario's involving branes have been proposed (see
e.g. \cite{cosmo}). Though popular in the media, their status is
mainly phenomenological, as it is often difficult to embed them in
string theory. One of the reasons is that string theory does not
allow any freedom in the choice of brane tensions and/or
interbrane interactions. In addition, in string theory such models
often lead to singularities that are hard to study.

In the remainder, we discuss some attempts to obtain and/or
understand time dependent backgrounds in string theory.

\subsection{time-dependent orbifolds}

One of the nice features of string theory is that it can deal with
certain types of singularities. In particular, if one starts with
a smooth string theory background and then makes some discrete
identifications, a procedure known as orbifolding, the resulting
singularities are well-understood and usually completely under
control. This leads to the question whether there any orbifold
constructions in string theory that provide a good toy model of
cosmological singularities, and/or of time-dependent backgrounds.
Many examples of such orbifolds have been studied recently
\cite{orbifoldlist}.

One of the simplest orbifolds one can think of is made by starting
with flat space. This has Poincar\'e invariance, so one can try to
make discrete identifications with respect to some finite subgroup
of the Poincar\'e group. For example, one can make an identifation
under a translation in a given direction, in which case that
direction becomes compact and turns into a circle. It is more
interesting to consider orbifolds that also involve the time
direction in some non-trivial way, so that the orbifold theory
becomes time-dependent. To some extent, the physics of such
time-dependent orbifolds can be extracted from the ambient theory
in flat space. However, the lack of invariance under time
translations, the non-existence of a good Wick rotation (and
therefore of a $+i\epsilon$ prescription in propagators), and the
absence of a Hamiltonian that is bounded from below make the
interpretation of these time dependent orbifolds rather confusing.
In addition, in case the group of identifications is not finite,
it has been argued that the resulting orbifold backgrounds are
generically unstable \cite{unstable}. This happens roughly because
a single particle in the orbifolded background corresponds to
infinitely many particles in the unorbifolded space, that are all
mapped into each other under the discrete identifications. This
infinite set of particles in the unorbifolded space tend to form a
black hole. To evade this, one needs a large number of non-compact
transversal directions, which makes the models much less
realistic.

\subsection{S-branes}

The D-branes we discussed in section~(\ref{branesec}) can move in
time, and are often stationary. One may wonder whether it is
possible to make extended objects that are localized in time, and
use these to generate interesting time-dependent backgrounds in
string theory. Such objects are not the same as instantons. Though
instantons are thought of as objects localized in time, they are
solutions of the Euclidean theory, not of the Minkowski theory,
and here we want solutions of the Minkowski equations of motion.
Since all directions along such localized objects are space-like,
they have been called S-branes. Some interesting recent work in
this direction has been done \cite{sbrane}, but at present, no
sufficiently stable S-branes have been found.

\subsection{more speculative ideas}

As we review below, it is difficult to obtain a well-controlled
solution of string theory with a positive cosmological constant.
This has led to the suggestion that perhaps we should modify
string theory in a more drastic way. For example, perhaps we
should start to think about non-local string theories. Such
theories have been discussed in \cite{nonlocal} as a useful
framework to discuss particle creation in string theory. Another
idea is to make sense of a version of string theory where some of
the fields are allowed to take imaginary values \cite{hull,ds2}.
Such theories can much more easily accommodate solutions with a
positive cosmological constant than ordinary string theories.

\subsection{de Sitter space}

Recall from section~4 that Anti-de Sitter space played a crucial
role in the AdS/CFT duality. It was a maximally symmetric solution
of the Einstein equations with a negative cosmological constant.
Similarly, there exists something called de Sitter space, which is
a maximally symmetric solution of the Einstein equations with a
positive cosmological constant. The metric of de Sitter space is
of the form $ds^2=-dt^2 + \cosh^2 t d\Omega_3^2$, in other words
it describes a three-sphere that has its minimum size at $t=0$ and
expands exponentially in the future and in the past.

During inflation the universe expanded exponentially, and was
approximately described by a de Sitter space. At this very moment
the universe again appears to be entering a de-Sitter phase,
driven by the small but nonzero cosmological constant that has
been observed. It is therefore natural to wonder whether there is
a solution of string theory that somehow involves de Sitter space,
or more generally, any solution of the Einstein equations with a
positive cosmological constant.

This is remarkably difficult to achieve\footnote{Supergravity
solutions of this type are e.g. discussed in
\cite{hull,dssugra}.}. Solutions with a negative cosmological
constant are easily generated by turning on field strengths for
some of the tensor fields of string theory, but this never leads
to a positive cosmological constant. There are no-go theorems
\cite{nogo} that state that there is no smooth solution of
supergravity that involves de Sitter space and a compact internal
space. These no-go theorems do not necessarily apply to string
theory to which supergravity is only a low-energy approximation.
Indeed, the no-go theorems assume a certain positive energy
condition which is violated in string theory: string theory has
negative tension brane-like objects (so-called orientifolds),
which could be crucial in obtaining solutions with a positive
cosmological constant; see \cite{stromeva} for a recent attempt
involving non-critical strings\footnote{Other attempts to obtain
interesting time-dependent backgrounds from non-critical string
theory can for example be found in \cite{mavro}.}. Unfortunately,
a well-controlled solution of string theory with a positive
cosmological constant remains out of reach for now.

Still, if one assumes that de Sitter solutions of string theory
exist, one may ask whether they would admit a dual field theory
description like we have in the case of the AdS/CFT
correspondence. Such a dual description would be a very powerful
tool in analyzing the physics of de Sitter space. Preliminary
results in this direction indicate that if anything, de Sitter
space should be dual to rather peculiar conformal field theories
\cite{ds2,dsholography}. This leads to an interesting
picture. In the case of the AdS/CFT correspondence, the extra
radial dimension of AdS had the interpretation of an energy scale
in the dual field theory. In de Sitter space, the role of the
radial coordinate of AdS is taken over by time, and time should
therefore somehow be related to an energy scale in the dual field
theory. This leads to the speculation that the transition in the
universe from the inflationary de Sitter phase to the present de
Sitter phase is described by some sort of renormalization group
flow in the putative dual field theory \cite{inflation}. If true,
this would provide a completely new way to think about the time
evolution of the universe.

\subsection{the inflaton as a tachyon}

Tachyons are particles with negative mass squared, and as free
particles they are unphysical. In interacting theories they do not
have to be unphysical at all. They often simply indicate that one
is not expanding around the true vacuum of the theory, which is
e.g. what happens in the standard model when expanding around zero
Higgs field. If we start with a theory not in its vacuum state, it
will undergo some dynamical process. For instance, it can decay to
its ground state, while radiating all energy in the initial state
away to infinity.

In string theory, we frequently find tachyons in the spectrum.
Probably the best known example is the purely bosonic string that
lives in 26 dimensions. It has a tachyon in its spectrum and is
therefore often discarded as being inconsistent. Of course, it is
possible that the bosonic string has some runaway potential for
the tachyon with no minimum at all; but it is equally well
possible that the bosonic string does have a stable ground state,
it is simply beyond our present capabilities to determine what
such a ground state should be.

Luckily, there are many string theory setups with tachyons where
the tachyon is under a reasonable amount of control. A good
example is a system consisting of D-branes and anti D-branes. Anti
D-branes are extended objects with the opposite quantum numbers
compared to D-branes, and anti D-branes and D-branes can
annihilate each other. This instability is reflected in string
theory by the presence of a tachyon, and has been studied in great
detail \cite{ddbar}.

It is an interesting question whether we can find time dependent
solutions in string theory by creating an unstable initial state,
and letting it evolve in time. The time dependence is then
generated by the dynamics of the tachyon degrees of freedom. Some
time dependent solutions describing a tachyon rolling down a
potential have been found in \cite{sen}. The endpoint of the
process is a gas of excited string states. The dynamics of such
processes can be captured by a simple effective field theory with
action $S\sim \int e^{-aT} \sqrt{1+(\partial_{\mu} T)^2}$
\cite{tact}, where $T$ is the tachyon degree of freedom. It is,
however, hard to get realistic models in this way, since there are
no naturally small parameters in the theory and the slow-roll
conditions of inflation will typically not be satisfied. In
addition, the nature of the gas of string states that remains was
calculated in an approximation that will break down sooner or
later, because the closed string states that can also be radiated
away have not been precisely taken into account \cite{closedt}.

Alternatively, one may try to use the tachyon to describe the
reheating of the universe after inflation has ended. This is
however harder to make explicit in string theory. For some further
discussions, see \cite{shiut,gibbonst}.

\subsection{string theory signatures in cosmology}

Is there a possibility to see signatures of string theory in
cosmology? As we go back in time, there is a moment in the history
of the universe when the temperature of the universe was so high
that classical gravity is no longer a good approximation, and in
order to describe the universe before this time a theory of
quantum gravity is needed. In particular such a theory of quantum
gravity should explain the initial conditions for the various
fields in the universe. These initial conditions are reflected in
the spectrum of the cosmic microwave background radiation. Since
we have no precise string theory description of the early
universe, we can only estimate the magnitude string theory effects
would have on the cosmic microwave background. Naively, string
theory effects will affect the power spectrum by terms of order
$(H/M)^2$, where $H$ is the Hubble constant and $M$ the scale of
new physics. However, this assumes a standard choice of vacuum
state for the fields. If we drop this assumption the effects can
be larger, of the order of $(H/M)$. With an optimistic choice of
$M$, this could be at the treshold of observational limits. It
would be an absolutely tremendous achievement if an experimental
signature of string theory could be obtained in this way! It will
nevertheless be difficult to disentangle any such effects from the
data, probably the spectrum of power fluctuations will not be
enough and the spectrum of tensor fluctuations will also be
needed. Besides this, there is still a lively debate going on
whether non-standard choices of vacuum states are physically
acceptable or not; see e.g. \cite{alpha} for further discussion of
this issue.

\section{ACKNOWLEDGEMENTS}

I would like to thank the organizers of the International
Conference on High Energy Physics 2002 for the invitation to
present this lecture. I would also like to thank a large number of
colleagues for various bits of insight they provided, and in
particular I would like to thank Gary Shiu for some very useful
discussions.


\end{document}